\documentclass[journal]{IEEEtran}

\usepackage{xcolor,soul,framed} %,caption

\usepackage[pdftex]{graphicx}
\graphicspath{{../pdf/}{../jpeg/}}
\DeclareGraphicsExtensions{.pdf,.jpeg,.png}

\usepackage[cmex10]{amsmath}
\usepackage{siunitx}
\usepackage{graphicx}
\usepackage{multirow}
\usepackage{multicol}
\usepackage{amssymb} 
\usepackage{amsmath}
\usepackage{mathtools}
\usepackage{cite}
\usepackage{siunitx}
\usepackage{cuted}
\usepackage{flushend}
\usepackage{multibib}
\usepackage{relsize}
\usepackage{stix}
\usepackage{balance}
\DeclarePairedDelimiterXPP\BigOSI[2]{\mathcal{O}}{(}{)}{}{\SI{#1}{#2}}

\begin{document}
\bstctlcite{IEEEexample:BSTcontrol}
    \title{Single-polarization Hybrid Hollow-core Anti-resonant Fiber Designs at 2 $\mu$m}
    
 \author{
           Herschel Herring, Md. Selim Habib,~\IEEEmembership{Senior Member,~IEEE,} ~\IEEEmembership{Senior Member, Optica}
      
  \thanks{Manuscript received March XX, 2023. M. Selim Habib acknowledges partial support of this research was provided by the Woodrow W. Everett, Jr. SCEEE Development Fund in cooperation with the Southeastern Association of Electrical Engineering Department Heads.}
   \thanks{H. Herring and M. Selim Habib are with the Department of Electrical and Computer Engineering, Florida Polytechnic University, FL-33805, USA (e-mail: mhabib@floridapoly.edu).}
  
 % <-this % stops a space
  }

% The paper headers
%\markboth{IEEE Photonic Technolgoy Letters, VOL.~XXX, NO.~XXX, June~2021}
%{Habib \MakeLowercase{\textit{et al.}}:XXX}
% title
\maketitle

% Abstract

\begin{abstract}
%\boldmath
 In this letter, to the best of our knowledge, a new type of hollow-core anti-resonant fiber (HC-ARF) design using hybrid silica/chalcogenide cladding is presented for single-polarization, high-birefringence, and endlessly single-mode operation at 2 $\mu$m wavelength. We show that the inclusion of a chalcogenide layer in the cladding allows strong suppression of $x$-polarization, while maintaining low propagation loss and single-mode propagation for $y$-polarization. The optimized HC-ARF design includes a combination of low propagation loss, high-birefringence, and polarization-extinction ratio (PER) or loss ratio of 0.02 dB/m, 1.2$\times$10$^{-4}$, >550 respectively, while the loss of the $x$-polarization is >20 dB/m. The proposed fiber may also be coiled to small bend radii while maintaining low bend-loss of $\approx$ 0.01--0.1 dB/m, and can potentially be used as polarization filter based on the different gap separations and bend conditions.
\end{abstract}


\IEEEpeerreviewmaketitle

% I. INTRODUCTION 
\section{Introduction}
%\IEEEPARstart{T}{he} 
\IEEEPARstart{T}{he} 2 $\mu$m wavelength is of particular interest in the field of optics and photonics due to its unique property of high absorption in water. This characteristic allows for an enormous range of optical fiber system applications such as biomarker detection, environmental observation, and molecular spectroscopy \cite{Scholle10}, \cite{6879250}. High light transmission, excellent polarization control, and high-birefringence are much needed for many optical devices 
%needed in these applications 
such as fiber sensors \cite{chen2008local}, fiber lasers \cite{lin1990polarisation}, optical amplifiers \cite{peng2007fundamental}, and fiber based gyroscopes \cite{terrel2011resonant}. %strong transmission performance and the ability to maintain control over the polarization state of light are necessary.
Although several designs in the past have achieved high-birefringence and single-polarization in solid-core \cite{hosaka1981low} and photonic band gap \cite{fini2014polarization} fibers, hollow-core anti-resonant fibers (HC-ARFs) have proven to be the foremost choice because of their unique and exceptional optical guidance in air \cite{sakr2020hollow}. HC-ARFs guide light through inhibited-coupling between the core and cladding modes (CMs) and anti-resonant effect \cite{couny2007generation}. This mechanism provides wide transmission bandwidth \cite{poletti2014nested,habib2015low,habib2016low,adamu2019deep}, extremely low power overlap with cladding tubes \cite{michieletto2016hollow}, low anomalous dispersion \cite{sakr2020interband} and extremely low loss \cite{jasion20220}. 

Achieving high-birefringence and single-polarization using the HC-ARF structure is a relatively new concept, which was first proposed in 2016 \cite{mousavi2016broadband}. Recently, a few high-bireringence and single-polarization HC-ARF designs have been proposed in the near-IR regime mostly at 1550 nm or 1064 nm with different cladding arrangements and number of layers in the cladding structure \cite{mousavi2016broadband,yan2020new,habib2021enhanced, hong2022highly}. High-birefringence and single-polarization were achieved either by using multiple nested resonators \cite{mousavi2016broadband,hong2022highly} or high index materials in the cladding \cite{yan2020new,habib2021enhanced}. Most recently, a four-fold \cite{hong2022highly} and six-fold \cite{yerolatsitis2020birefringent} symmetry bi-thickness HC-ARF were experimentally reported with outstanding optical performances. For example, in \cite{hong2022highly}, a phase birefringence of $2.35\times 10^{-5}$ and $9.1\times 10^{-5}$ at 1550 and 1589 nm were achieved respectively.

However, to the best of our knowledge, single-polarization, and single-mode HC-ARFs have not yet been investigated at 2 $\mu$m. In this paper, we propose a semi-nested HC-ARF design that utilizes a hybrid silica/chalchogenide cladding to attain low propagation loss, single-polarization, high-birefringence, and endlessly single-mode operation simultaneously at 2 $\mu$m.

% figure-1
\begin{figure*}[t!]
  \begin{center}
  \includegraphics[width=7in]{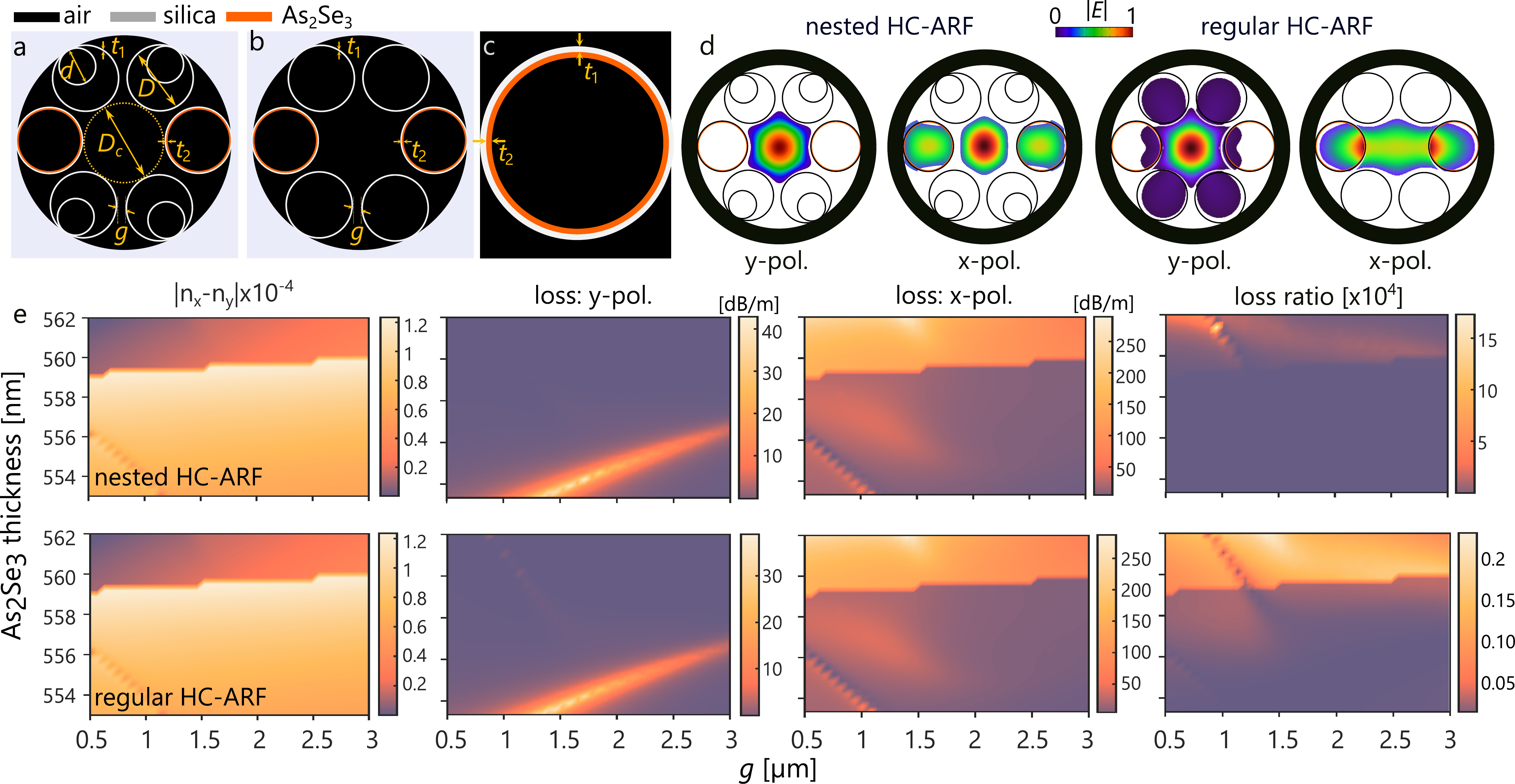}\\
   \caption{HC-ARF geometries:  6-tube (a)  semi-nested HC-ARF with six nested tubes and (b) regular HC-ARF in which the nested tubes are not present in the outer tubes. (c) Both fibers have a core diameter of $D_\text{c} = 56$ $\mu$m, t$_{1}$/t$_{2}$ = wall thickness of silica/$\text{As}_2\text{Se}_3$ tubes, $d$ = nested tube diameter, $D$ = outer tube diameter, $g$ = separation between the outer tubes. (d) Normalized mode-field profiles of both polarizations for nested and regular HC-ARF at 2 $\mu$m. The color bar shows the intensity distributions in a linear scale. The mode-field profile ($y$-pol.) of nested HC-ARF is more confined to the core compared to the regular HC-ARF. (e) The FE- simulated results of birefringence: $|n_x-n_y|$, propagation loss in dB/m of $y$- and $x$- polarizations, and PER or loss ratio of both polarizations. Top panel: nested HC-ARF, and bottom panel: regular HC-ARF. The simulations were perfomed at 2 $\mu$m. To plot the 2D surface plots, gap separation, $g$, and $\text{As}_2\text{Se}_3$ wall thickness, $t_2$ were scanned with 30 and 40 data points respectively and between the data points are interpolated.}  \label{fig:fig_1}
  \end{center}
\end{figure*}

% figure-2
\begin{figure*}
  \begin{center}
  \includegraphics[width=7in]{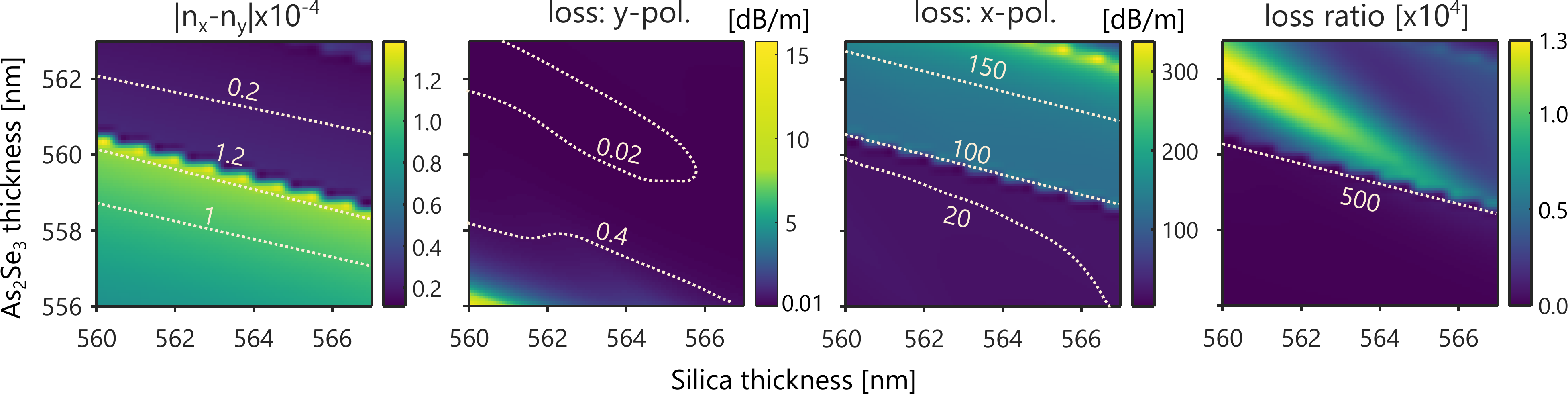}\\
    \caption{
    The FE- simulated results of birefringence: $|n_x-n_y|$, propagation loss of $y$- and $x$- polarizations, and loss ratio of both polarizations as a function of $\text{As}_2\text{Se}_3$ thickness, $t_{2}$ with different values of silica thickness, $t_{1}$. The HC-ARF has core diameter, $D_\text{c}$ = 56 $\mu$m, normalized tube ratio, $d/D$ = 0.5, and gap separation, $g$ = 2 $\mu$m. The simulations were performed at 2 $\mu$m. Silica wall thickness, $t_1$, and $\text{As}_2\text{Se}_3$ wall thickness, $t_2$ were scanned with 25 and 30 data points respectively and between the data points are interpolated.
    }\label{fig:fig_2}
  \end{center}
\end{figure*}

% Section II
\section{Fiber Geometry}
The HC-ARF geometries investigated in this work are displayed in Fig.~\ref{fig:fig_1}(a--b). Fig.~\ref{fig:fig_1}(a) shows a semi-nested 6-tube non-touching HC-ARF architecture in which chalcogenide ($\text{As}_2\text{Se}_3$) tubes (orange color) are inserted into two horizontally placed silica tubes and two of the nested tubes are removed from those silica tubes to ensure low loss in one polarization state ($y$-polarization) and high loss for the other polarization state ($x$-polarization). Chalcogenide is used due to its high transmission at 2 $\mu$m. %The refractive index at a specific wavelength for can be calculated by the Sellmeier equation: $\center n(\lambda) = \sqrt{1+\lambda^{2}(\frac{A_{0}^{2}}{\lambda^{2}-A_{1}^{2}}+\frac{A_{0}^{2}}{\lambda^{2}-19^{2}}+\frac{A_{0}^{2}}{\lambda^{2}-4A_{1}^{2}})}$. 
The HC-ARF has core diameter of $D_\text{c}$, tube diameter $D$, wall thickness of silica/$\text{As}_2\text{Se}_3$ tubes, $t_{1}/t_2$, nested tube ratio, $d/D$, and a gap separation between the outer tubes, $g$. %The proposed HC-ARF architecture and fiber  parameters ensure single-polarization, high-birefringence, endlessly single-mode operation simultaneously at 2 $\mu$m.
Fig.~\ref{fig:fig_1}(b) shows a regular HC-ARF design without nested tubes. In all our simulations, we choose a relatively large core diameter of $D_\text{c}$ = 56 $\mu$m to ensure low loss in the $y$-polarization. However, we optimize the silica/chalcogenide tube thickness ($t_1/t_2$), and gap separation, $g$ to ensure single-polarization, single-mode, and high-birefringence. The outer diameter $D$ is related to the core diameter $D_\text{c}$, wall thickness $t_1$, and number to tubes $N$, which can be written as \cite{wei2017negative}:
%$D=(\frac{D_\text{c}}{2}sin(\frac{\pi}{N})-\frac{g}{2}-t{_1}(1-sin(\frac{\pi}{N})))(1-sin(\frac{\pi}{N}))^{-1}$.  
\begin{equation}
\begin{split}
%D=\frac{D_\text{c}}{2}sin(\frac{\pi}{N})-\frac{g}{2}-t{_1}(1-sin(\frac{\pi}{N})),
\frac{D}{2} = \frac{\frac{D_\text{c}}{2}\text{sin}(\frac{\pi}{N})-\frac{g}{2}-t{_1}(1-\text{sin}(\frac{\pi}{N}))}{1-\text{\text{sin}}(\frac{\pi}{N})}.
\end{split}
\label{eq:eq1}
\end{equation}
%The gap separation, $g$ can be written as \cite{wei2017negative} $g=D_c sin(\pi/ N)-(D+2t)(1-sin(\pi/N))$, where $D_c$ = core diameter, $D$ = outer tube diameter, $t$ = wall thickness of outer tubes, and $N$ is the number of tubes.

%In our all simulations, a small penetration of $t_1/2$ of all cladding tubes into the outer silica tube was considered which is the typical case in fabricated fibers \cite{belardi2014hollow}.

% Numerical Results and Discussion Section
\section{Numerical Results and Discussion}
The simulations were performed using fully-vectorial finite-element (FE) modeling. 
%based on the \textsc{Comsol}$^{\circledR}$--MATLAB Livelink software package. 
A perfectly-matched layer (PML) boundary was placed outside the fiber domain to accurately calculate the confinement loss.
The mesh size and PML boundary conditions were chosen similar to \cite{poletti2014nested,habib2019single,habib2021enhanced}. The propagation loss was calculated by considering confinement loss, effective material loss \cite{habib2019single}, and surface scattering loss (SSL) \cite{roberts2005ultimate,poletti2014nested}. %The calculated power overlap of silica and $\text{As}_2\text{Se}_3$ are <$1.5\times10^{-4}$ and <$2.5\times10^{-5}$ respectively at 2 $\mu$m.

% Optimization of g and t2 subsection
\subsection{Optimization of gap separation and  $\text{As}_2\text{Se}_3$ wall  thickness}

We started our investigations towards achieving high-birefringence, single-mode, and single-polarization by optimizing the gap separation, $g$, and the chalcogenide ($\text{As}_2\text{Se}_3$) wall thickness, $t_2$ for a fixed the core diameter, $D_\text{c}$ = 56 $\mu$m, silica wall thickness, $t_1$ = 560 nm, and normalized nested tube ratio, $d/D$ = 0.5. The desired properties of low loss, single-polarization, and high-birefringence are highly dependent on these two parameters. In particular, the effect of $g$ is critical to examine since it heavily influences how the fundamental-mode (FM) of one polarization is guided in the core while that of the other polarization spreads to the cladding. Fig.~\ref{fig:fig_1}(d) shows the mode-field profiles of both nested and regular HC-ARFs in which it can be seen that the light of $y$-polarization is almost entirely guided in the core, while the light of $x$-polarization spreads more towards the $\text{As}_2\text{Se}_3$ cladding. It is also clear that there is enhanced performance with the nested HC-ARF compared to regular HC-ARF as more of the light is well guided in the core and weakly leaks towards the cladding. 

The FE-simulated results for semi-nested and regular HC-ARF are shown in Fig.~\ref{fig:fig_1}(e). From the 2D surface plots, we can see that there is low loss for $y$-polarization, high loss for $x$-polarization, high-birefringence, and high loss ratio around the region where $g$ = 2 $\mu$m and $t_{2}$ = 559 nm. The graphical patterns for the semi-nested and regular HC-ARF are similar in regards to birefrigence and the loss of both polarizations, however the semi-nested structure displays a much higher loss ratio. Numerically, the semi-nested structure also shows significantly lower loss of the $y$-polarization with a minimum FM loss of $\approx$ 0.0011 dB/m comparing to that of the regular structure at $\approx$ 0.05 dB/m. It is also evident that the ideal range is more highly impacted by the value of the $\text{As}_2\text{Se}_3$ wall thickness, since there is a visible boundary near $t_2$ = 559 nm where the birefringence drops greatly from >$1.23\times10^{-4}$ to <$0.27\times10^{-4}$. For both semi-nested and regular HC-ARFs, the maximum possible loss of $x$-polarization, birefringence, and loss ratio as well as the minimum loss of the $y$-polarization do not all occur at the same values of $g$ and $t_{2}$. For example, the semi-nested structure's best values are 290 dB/m, >$1.23\times10^{-4}$, >$1.7\times10^{5}$, and $\approx$ 1 dB/km respectively, and all occurring for different values of $g$ and $t_{2}$. The optimal region where high-birefringence of >$1.0\times10^{-4}$ is achieved while maintaining an FM loss of the $y$-polarization of <$0.05$ dB/m is highly sensitive, and occurs where $g$ = 2 $\mu$m and 559 nm < $t_{2}$ < 559.5 nm. A polarization-extinction ratio (PER) or loss ratio of >$550$ is shown in this region as well. Values near this $t_{2}$ range were chosen for analysis in further sections with the anticipation that high-birefringence, low loss, and single-mode operation could be enhanced with additional parameter tuning.

% figure-3
\begin{figure}
  \begin{center}
  \includegraphics[width=3.4in]{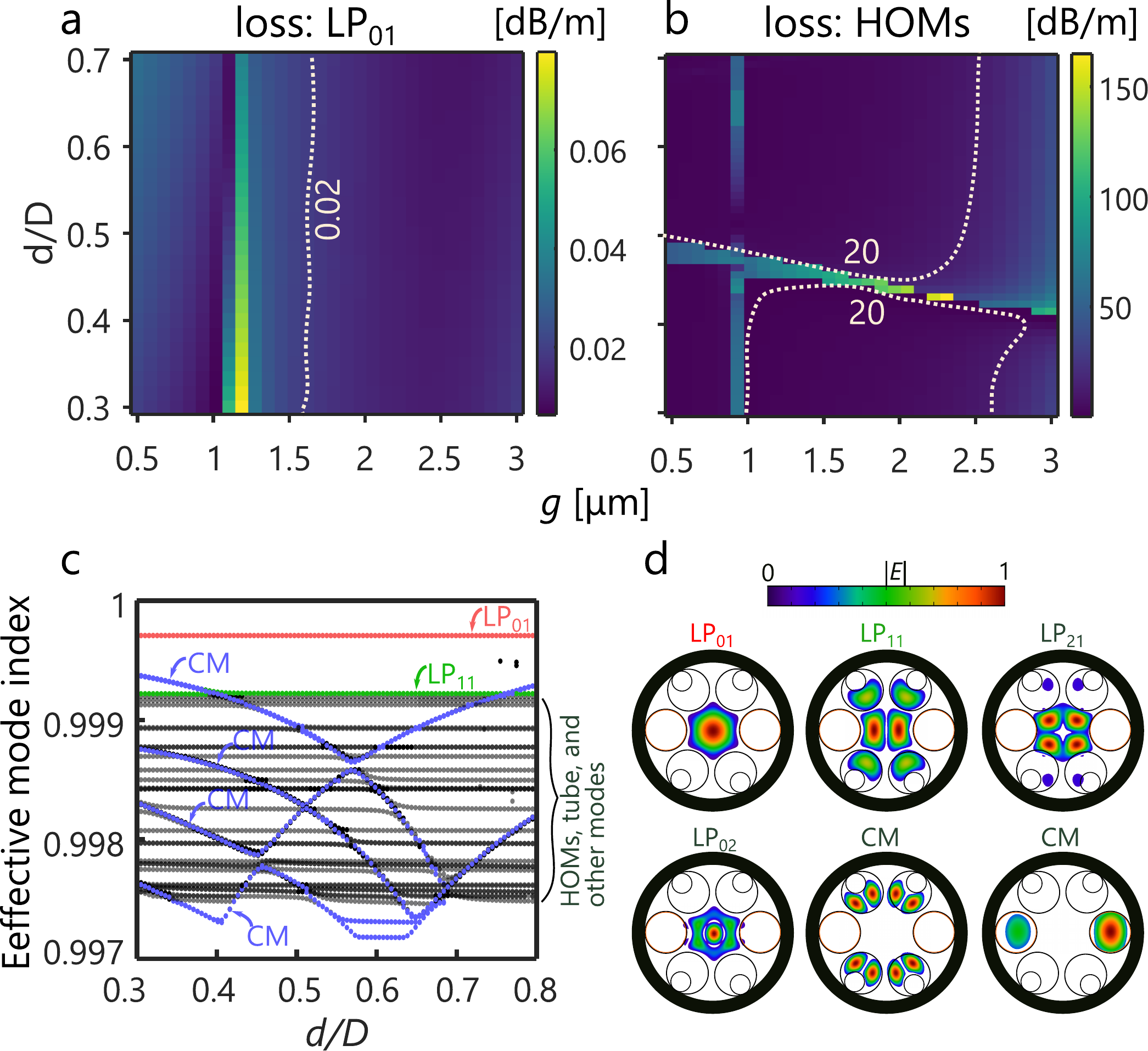}\\
    \caption{
    The FE- simulated propagation loss of $y$-polarization for (a)  $LP_\text{01}$-mode and (b) HOMs for varying values of $d/D$ and gap separation, $g$.  Results were scanned with 50 and 30 data points respectively and between the data points are interpolated. Effect of changing normalized nested tube ratio, $d/D$ on (c) Effective mode index with a core diameter, $D_\text{c}$ = 56 $\mu$m, $t_1/t_2$ = 564/559 nm, and gap separation, $g$ = 2 $\mu$m. The normalized tube ratio, $d/D$ were scanned from 0.3 to 0.8 by simulating 50 modes at 2 $\mu$m. (d) The electric field intensities of the first four core-guided modes and CMs are shown for $d/D$ $\approx$0.42 and $g$ = 2 $\mu$m on a linear color scale.
    }\label{fig:fig_3}
  \end{center}
\end{figure}

%The first two parameters that we optimized in our analysis were the gap separation, $g$, and the chalcogenide ($\text{As}_2\text{Se}_3$) wall thickness. 
%Here, we fixed the core diameter $D_\text{c}$ = 56 $\mu$m, silica wall thickness, $t_1$ = 560 nm, and normalized nested tube ratio, $d/D$ = 0.5. 
%The desired properties of low loss, single-polarization, and high-birefringence are highly dependent on the gap separation $g$ at varying wall thicknesses for both semi-nested and regular fiber geometries. These results can be seen in Fig.~\ref{fig:fig_1}. 

\subsection{Optimization of silica/$\text{As}_2\text{Se}_3$ wall thickness ($t_{1}$/$t_{2}$)}

In this section, we optimized the wall thicknesses of silica/$\text{As}_2\text{Se}_3$, while fixing our gap separation, $g$ = 2 $\mu$m as found in the previous section. Here, we maintained the fixed core diameter and normalized nested tube ratio of $D_\text{c}$ = 56 $\mu$m and $d/D$ = 0.5 respectively. From the results of the 2D contour plots seen in Fig.~\ref{fig:fig_2}, we can see that the wall thicknesses are crucial to achieving low loss, high-birefringence, and single-polarization. As mentioned previously, we chose the value of $g$ = 2 $\mu$m with the anticipation that further parameter tuning could lead to enhanced birefringence. From the simulated results of the silica/$\text{As}_2\text{Se}_3$ sweep, high-birefringence of >$1.20\times10^{-4}$ was achieved in the region where $t_{1}$ = 564 nm and $t_{2}$ = 559 nm. At these values, the FM loss of the $y$-polarization can be made as low as $0.02$ dB/m while the loss of the $x$-polarization is >$20$ dB/m. The loss ratio is in the range of 578-970 here as well. Again, the birefringence is highly sensitive to changes in the wall thickness of $\text{As}_2\text{Se}_3$ and a sharp drop can be seen once the it exceeds 559 nm. These results demonstrate that by sufficient tuning of the silica thickness $t_{1}$, we are able to enhance the properties of low loss, high-birefringence, and single-polarization. By doing so, we were able to increase high-birefringence from >$1.0\times10^{-4}$ to >$1.2\times10^{-4}$ and loss ratio from >400 to >550 while reducing the FM loss of the $y$-polarization from <$0.05$ dB/m to <$0.03$ dB/m.

\subsection{Single-mode operation and higher-order mode suppression}

Following the examination of the low loss, high-birefringence, and single-polarization, the effectively single-mode operation and the suppression of the higher-order modes (HOMs) is demonstrated by optimization of the gap separation, $g$, and the nested tube ratio, $d/D$. The FE- simulated results for this are shown in Fig.~\ref{fig:fig_3}, where we can easily see that there is a large region where the loss of the FM is very low (<0.02 dB/m) while the loss of the HOMs is high (>20 dB/m). It is also important to note that FM loss is more heavily influenced by changes in $g$ compared to changes in $d/D$. However, $d/D$ has a significant impact on the loss of the HOMs. The region of highest loss of the HOMs becomes narrow for the range 1 $\mu$m < $g$ < 2.5 $\mu$m, where the value of $d/D$ must be precise in order to obtain effectively single-mode operation. The coupling between the core-guided HOMs and CMs is shown in Fig.~\ref{fig:fig_3}(c--d). The effective mode index of the FM-like mode (red dotted line) remains unchanged as a function of $d/D$, avoiding any phase matching with CMs. This mechanism confirms low loss guidance. However, the effective index of HOMs strongly couple with CMs and tube modes at various values of $d/D$ (also see Fig.~\ref{fig:fig_3}(d), ensuring high loss for HOMs, thus confirming effective single-mode operation. 

%\subsection{Core diameter optimization}

%Next, we investigated the effect of changing the core diameter $D_\text{c}$ on the FM loss of both polarizations for a fixed $g$ = 2 $\mu$m, silica wall thickness,  $t_1/t_2$ = 564/559 nm, and normalized nested tube ratio, $d/D$ = 0.5. The results of the core diameter analysis revealed that the loss of the $y$-polarization remains low and relatively unchanged for all values while the loss of the $x$-polarization is significantly higher yet decreasing with the core diameter. For our previous analyses, we fixed the core diameter $D_\text{c}$ = 56 $\mu$m, however saw that the FM of the $x$-polarization can be made more lossy for lower core diameter values around 50 -- 52 $\mu$m. There is strong high-birefringence in this region of >$1.48\times10^{-4}$, however the loss of the $y$-polarization increases to >$0.1$ dB/m at $D_\text{c}$ = 50 $\mu$m specifically. 

%Since high-birefringence of >$1.3\times10^{-4}$ is achieved for any value of $D_\text{c}$, an optimal region can be determined simply by where the loss of the $y$-polarization is the lowest. This occurs in the range of 55.6 $\mu$m  < $D_\text{c}$ < 56.8 $\mu$m, where the FM loss of the $y$-polarization remains <$0.2$ dB/m. Thus we have shown that by changing the core diameter $D_\text{c}$, we can achieve even better performance in regards to low loss, high-birefringence, and single-polarization.

\subsection{Bend loss analysis}

\begin{figure}
  \begin{center}
  \includegraphics[width=3.35in]{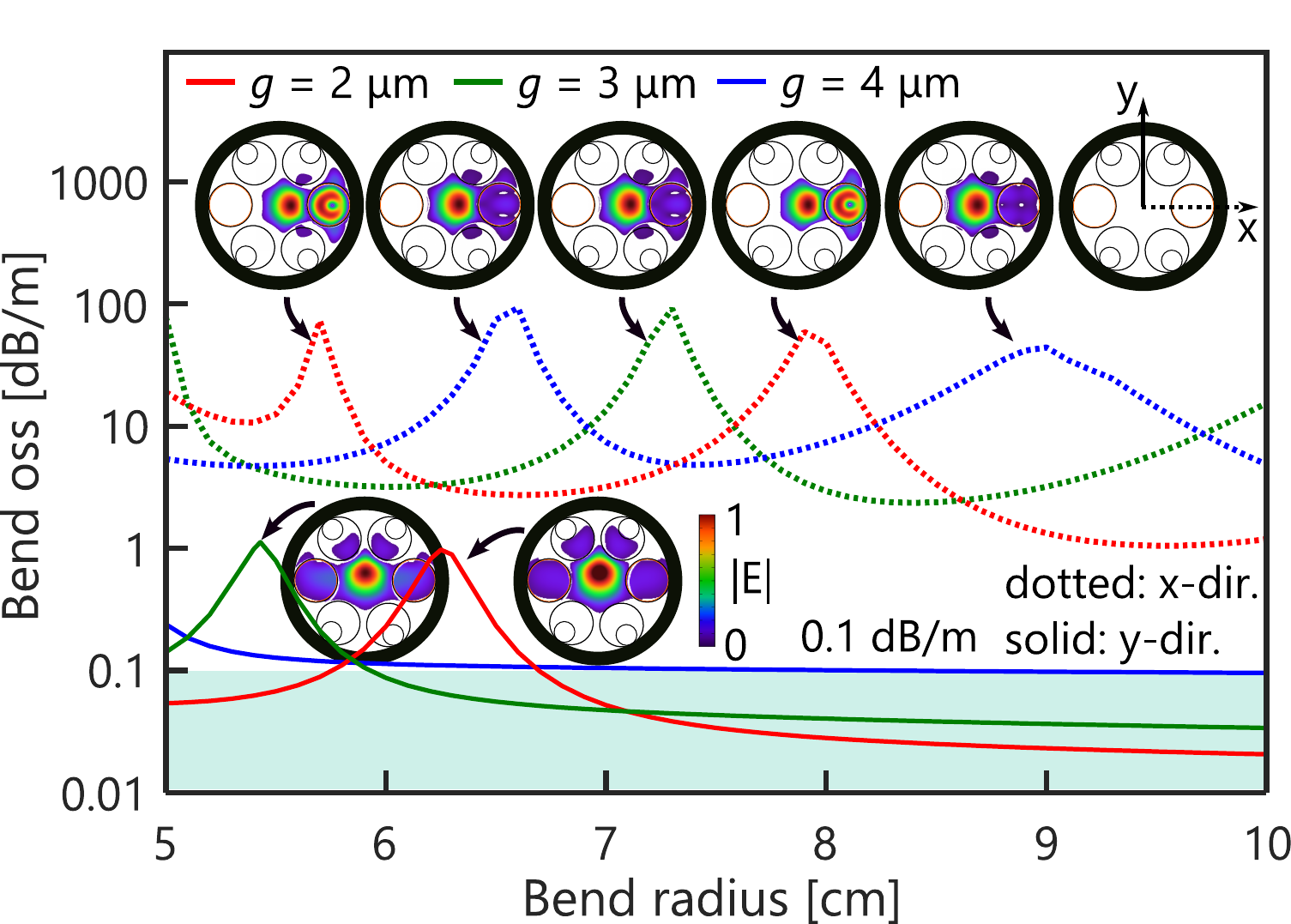}\\
    \caption{
    Calculated bend loss of $y$-polarization vs. bend radius from 5 cm to 10 cm in steps of 0.1 cm for different gap separations for $x$- and $y$-bend directions. The HC-ARF has core diameter, $D_\text{c}$ = 56 $\mu$m, silica/$\text{As}_2\text{Se}_3$wall thickness, $t_1/t_2$ = 564/559 nm, gap separation, $d/D$ = 0.65, and $g$ = 2 $\mu$m (red), 3 $\mu$m (green), and 4 $\mu$m (blue). The simulations were performed at 2 $\mu$m. The high loss peaks can be used for polarization filters. Inset: the electric field intensities on a linear color scale are shown for different bend radii.
    }\label{fig:fig_4}
  \end{center}
\end{figure}

Lastly, we examined the bend loss using \cite{poletti2014nested} of
semi-nested HC-ARF for a range of bend radii $R_\text{b}$ of 5 -- 10 cm for different $g$ = \{2, 3, 4\} $\mu$m, fixed core diameter $D_\text{c}$ = 56 $\mu$m, $t_1/t_2$ = 564/559 nm, and normalized nested tube ratio, $d/D$ = 0.65. The bend loss of $x$-direction is much higher than $y$-direction and has strong light coupling to the tube modes as expected due to the missing nested tubes in the $x$-directions. The strong coupling between the core modes and tubes modes occur at different bend radii and the coupling shifts for different $g$ values. This phenomenon is interesting because it indicates that the fiber can be used as a polarization filter at those gap separations and bend conditions as displayed in Fig.~\ref{fig:fig_4}.     

% figure-4

% Conclusion
\section{Conclusion}
In conclusion, we present a novel HC-ARF architecture that has been developed and validated to have single-polarization, high-birefringence, and single-mode operation characteristics at the 2 $\mu$m wavelength. Through extensive fiber parameter optimizations based on regular and semi-nested hybrid silica/chalcogenide HC-ARF structures, we demonstrated that the semi-nested geometry has better performance compared to regular HC-ARF. We also found a favorable range of the gap separation and chalcogenide wall thickness. From there, we were able to tune the silica thickness to obtain enhanced low loss, high-birefringence, and polarization-extinction ratio of 0.02 dB/m, 1.2$\times$10$^{-4}$, and >550 respectively. Furthermore, effectively single-mode operation was achieved by suitably choosing the nested tube ratio, $d/D$. We found that the loss of the HOMs could be made as high as 20 dB/m while the loss of the FM was <0.02 dB/m for a large range of $g$ and $d/D$ values. Finally, we investigated the effect of changing the bend radius in both $x$ and $y$-directions. %We found that FM bend loss for $x$-direction was much higher than the $y$-direction due to strong light coupling to the tube modes. 
The strong coupling between the core-guided modes and CMs lead to an interesting phenomenon through which the proposed fiber can be used as polarization filter. The exceptional polarization, low propagation loss, and single-mode properties presented in this work will pave the way toward practical applications at the highly appealing 2 $\mu$m wavelength. 

\section*{Acknowledgment}
The authors would like to thank Dr. Rodrigo Amezcua-Correa for providing high performance computational support.

%The author would like to thank Dr. Mohammed Saleh for useful discussions.

\bibliographystyle{IEEEtran}
\bibliography{IEEEabrv, Bibliography}

\begin{thebibliography}{10}
\providecommand{\url}[1]{#1}
\csname url@rmstyle\endcsname
\providecommand{\newblock}{\relax}
\providecommand{\bibinfo}[2]{#2}
\providecommand\BIBentrySTDinterwordspacing{\spaceskip=0pt\relax}
\providecommand\BIBentryALTinterwordstretchfactor{4}
\providecommand\BIBentryALTinterwordspacing{\spaceskip=\fontdimen2\font plus
\BIBentryALTinterwordstretchfactor\fontdimen3\font minus
  \fontdimen4\font\relax}
\providecommand\BIBforeignlanguage[2]{{%
\expandafter\ifx\csname l@#1\endcsname\relax
\typeout{** WARNING: IEEEtran.bst: No hyphenation pattern has been}%
\typeout{** loaded for the language `#1'. Using the pattern for}%
\typeout{** the default language instead.}%
\else
\language=\csname l@#1\endcsname
\fi
#2}}

\bibitem{Scholle10}
K.~Scholle, S.~Lamrini, P.~Koopmann, and P.~Fuhrberg, ``2 $\mu$m laser sources
  and their possible applications,'' in \emph{Frontiers in Guided Wave Optics
  and Optoelectronics}.\hskip 1em plus 0.5em minus 0.4em\relax Rijeka:
  IntechOpen, 2010.

\bibitem{6879250}
S.~B. Mirov, \emph{et~al.}, ``Progress in mid-{IR} lasers based on {C}r and
  {F}e-doped ii–vi chalcogenides,'' \emph{IEEE Journal of Selected Topics in
  Quantum Electronics}, vol.~21, no.~1, pp. 292--310, 2015.

\bibitem{chen2008local}
S.~Chen, \emph{et~al.}, ``Local electric field enhancement and polarization
  effects in a surface-enhanced raman scattering fiber sensor with chessboard
  nanostructure,'' \emph{Optics Express}, vol.~16, no.~17, pp.
  13\,016--13\,023, 2008.

\bibitem{lin1990polarisation}
J.~T. Lin and W.~A. Gambling, ``Polarisation effects in fibre lasers:
  phenomena, theory and applications,'' in \emph{IEE Colloquium on Polarisation
  Effects in Optical Switching and Routing Systems}.\hskip 1em plus 0.5em minus
  0.4em\relax IET, 1990, pp. 10--1.

\bibitem{peng2007fundamental}
X.~Peng and L.~Dong, ``Fundamental-mode operation in polarization-maintaining
  ytterbium-doped fiber with an effective area of 1400 $\mu$m$^2$,''
  \emph{Optics Letters}, vol.~32, no.~4, pp. 358--360, 2007.

\bibitem{terrel2011resonant}
M.~A. Terrel, M.~J. Digonnet, and S.~Fan, ``Resonant fiber optic gyroscope
  using an air-core fiber,'' \emph{Journal of Lightwave Technology}, vol.~30,
  no.~7, pp. 931--937, 2011.

\bibitem{hosaka1981low}
T.~Hosaka, \emph{et~al.}, ``Low-loss single polarisation fibres with
  asymmetrical strain birefringence,'' \emph{Electronics Letters}, vol.~17,
  no.~15, pp. 530--531, 1981.

\bibitem{fini2014polarization}
J.~M. Fini, \emph{et~al.}, ``Polarization maintaining single-mode low-loss
  hollow-core fibres,'' \emph{Nature Communications}, vol.~5, no.~1, pp. 1--7,
  2014.

\bibitem{sakr2020hollow}
H.~Sakr, \emph{et~al.}, ``Hollow core optical fibres with comparable
  attenuation to silica fibres between 600 and 1100 nm,'' \emph{Nature
  communications}, vol.~11, no.~1, pp. 1--10, 2020.

\bibitem{couny2007generation}
F.~Couny, \emph{et~al.}, ``Generation and photonic guidance of multi-octave
  optical-frequency combs,'' \emph{Science}, vol. 318, no. 5853, pp.
  1118--1121, 2007.

\bibitem{poletti2014nested}
F.~Poletti, ``Nested antiresonant nodeless hollow core fiber,'' \emph{Optics
  Express}, vol.~22, no.~20, pp. 23\,807--23\,828, 2014.

\bibitem{habib2015low}
M.~S. Habib, O.~Bang, and M.~Bache, ``Low-loss hollow-core silica fibers with
  adjacent nested anti-resonant tubes,'' \emph{Optics Express}, vol.~23,
  no.~13, pp. 17\,394--17\,406, 2015.

\bibitem{habib2016low}
M.~S. Habib \emph{et~al.}, ``Low-loss single-mode hollow-core fiber with
  anisotropic anti-resonant elements,'' \emph{Optics Express}, vol.~24, no.~8,
  pp. 8429--8436, 2016.

\bibitem{adamu2019deep}
A.~I. Adamu, \emph{et~al.}, ``Deep-{UV} to mid-{IR} supercontinuum generation
  driven by mid-{I}{R} ultrashort pulses in a gas-filled hollow-core fiber,''
  \emph{Scientific Reports}, vol.~9, no.~1, pp. 1--9, 2019.

\bibitem{michieletto2016hollow}
M.~Michieletto, \emph{et~al.}, ``Hollow-core fibers for high power pulse
  delivery,'' \emph{Optics Express}, vol.~24, no.~7, pp. 7103--7119, 2016.

\bibitem{sakr2020interband}
H.~Sakr, \emph{et~al.}, ``Interband short reach data transmission in ultrawide
  bandwidth hollow core fiber,'' \emph{Journal of Lightwave Technology},
  vol.~38, no.~1, pp. 159--165, 2020.

\bibitem{jasion20220}
G.~T. Jasion, \emph{et~al.}, ``0.174 d{B}/km hollow core double nested
  antiresonant nodeless fiber (dnanf),'' in \emph{2022 Optical Fiber
  Communications Conference and Exhibition (OFC)}.\hskip 1em plus 0.5em minus
  0.4em\relax IEEE, 2022, pp. 1--3.

\bibitem{mousavi2016broadband}
S.~A. Mousavi, S.~R. Sandoghchi, D.~J. Richardson, and F.~Poletti, ``Broadband
  high birefringence and polarizing hollow core antiresonant fibers,''
  \emph{Optics Express}, vol.~24, no.~20, pp. 22\,943--22\,958, 2016.

\bibitem{yan2020new}
S.~Yan, \emph{et~al.}, ``A new method to achieve single-polarization guidance
  in hollow-core negative-curvature fibers,'' \emph{IEEE Access}, vol.~8, pp.
  53\,419--53\,426, 2020.

\bibitem{habib2021enhanced}
M.~S. Habib, A.~Adamu, C.~Markos, and R.~Amezcua-Correa, ``Enhanced
  birefringence in conventional and hybrid anti-resonant hollow-core fibers,''
  \emph{Optics Express}, vol.~29, no.~8, pp. 12\,516--12\,530, 2021.

\bibitem{hong2022highly}
Y.-f. Hong, \emph{et~al.}, ``Highly birefringent anti-resonant hollow-core
  fiber with a bi-thickness fourfold semi-tube structure,'' \emph{Laser \&
  Photonics Reviews}, vol.~16, no.~5, p. 2100365, 2022.

\bibitem{yerolatsitis2020birefringent}
S.~Yerolatsitis, \emph{et~al.}, ``Birefringent anti-resonant hollow-core
  fiber,'' \emph{Journal of Lightwave Technology}, vol.~38, no.~18, pp.
  5157--5162, 2020.

\bibitem{wei2017negative}
C.~Wei, R.~J. Weiblen, C.~R. Menyuk, and J.~Hu, ``Negative curvature fibers,''
  \emph{Advances in Optics and Photonics}, vol.~9, no.~3, pp. 504--561, 2017.

\bibitem{habib2019single}
M.~S. Habib, \emph{et~al.}, ``Single-mode, low loss hollow-core anti-resonant
  fiber designs,'' \emph{Optics Express}, vol.~27, no.~4, pp. 3824--3836, 2019.

\bibitem{roberts2005ultimate}
P.~Roberts, \emph{et~al.}, ``Ultimate low loss of hollow-core photonic crystal
  fibres,'' \emph{Optics Express}, vol.~13, no.~1, pp. 236--244, 2005.

\end{thebibliography}
\vfill
%\balance
\end{document}